# Estimating total cross sections from neutron-nucleus collisions using a simple functional form

Pradip Deb[*]

*School of Physical and Chemical Sciences, Queensland University of Technology, QLD 4001, Australia*

Ken Amos[†]

*School of Physics, The University of Melbourne, VIC 3010, Australia*

Steven Karataglidis[‡]

*Department of Physics and Electronics, Rhodes University, Grahamstown, South Africa*

**Abstract**

Total cross sections for neutron scattering from nuclei, with energies ranging from 10 to 600 MeV and from many nuclei spanning the mass range $^6$Li to $^{238}$U, have been analyzed using a simple, three-parameter, functional form. The calculated cross sections are compared with results obtained by using microscopic (*g*-folding) optical potentials as well as with experimental data. The functional form reproduces those total cross sections very well. When allowance is made for Ramsauer-like effects in the scattering, the parameters of the functional form required vary smoothly with energy and target mass. They too can be represented by functions of energy and mass.

## I. INTRODUCTION

Total cross sections and total reaction cross sections from the scattering of nucleons by nuclei and for energies to 600 MeV or more, are required in a number of fields of study in basic science as well as many of applied nature. Often, those cross sections have been evaluated using phenomenological optical potentials and much effort has gone into defining global sets of parameter values for those optical potentials with which to estimate cross sections as yet unmeasured. In a recent study, Koning and Delaroche (Koning and Delaroche, 2003) gave a detailed specification of such. However, it would be utilitarian if total and total reaction cross sections were well approximated by a simple convenient function form. Indeed that is so and we show herein that there is a simple three parameter function form one can use to form estimates without recourse to optical potential calculations. Further, the required values of the three parameters of that function form themselves trend sufficiently smoothly with energy and mass that they too may be represented by functional forms.

---

[*] Electronic Address: pkd@physics.unimelb.edu.au
[†] Electronic Address: amos@physics.unimelb.edu.au
[‡] Electronic Address: S.Karataglidis@ru.ac.za

## II. FORMALISM

The total cross sections for nucleon scattering from nuclei can be expressed in terms of partial wave scattering matrices specified at energies $E \propto k^2$, namely $S_l^{\pm}(k) = \eta_l^{\pm}(k) e^{2i\Re[\delta_l^{\pm}(k)]}$, where $\delta_l^{\pm}(k)$ are the (complex) scattering phase shifts and $\eta_l^{\pm}(k)$ are the moduli of the $S$ matrices. The superscript designates $j = l \pm 1/2$. In terms of these quantities, diverse partial wave dependent partial cross sections, $\sigma_l^{(X)}(k)$, can be formed whereby the elastic, reaction (absorption), and total cross sections respectively are given by

$$\sigma_{el}(E) = \frac{\pi}{k^2} \sum_l \sigma_l^{(el)}(k); \sigma_R(E) = \frac{\pi}{k^2} \sum_l \sigma_l^{(R)}(k);$$

$$\sigma_T(E) = \sigma_{el}(E) + \sigma_R(E) = \frac{2\pi}{k^2} \sum_l \sigma_l^{(T)}(k), \qquad (1)$$

where

$$\sigma_l^{(el)}(k) = (l+1)|S_l^+(k) - 1|^2 + l|S_l^-(k) - 1|^2;$$

$$\sigma_l^{(R)}(k) = (l+1)[1 - \eta_l^+(k)^2] + l[1 - \eta_l^-(k)^2];$$

$$\sigma_l^{(T)}(k) = (l+1)\{1 - \eta_l^+(k)\cos(2\Re[\delta_l^+(k)])\} + l\{1 - \eta_l^-(k)\cos(2\Re[\delta_l^-(k)])\}. \qquad (2)$$

Therein the $\sigma_l^{(X)}$ are defined as partial cross sections of the total elastic, total reaction, and total scattering itself. For proton scattering, because Coulomb amplitudes diverge at zero degree scattering, only total reaction cross sections are measured. Nonetheless study of such data (Amos et al. 2002, Deb and Amos, 2003) established that partial total reaction cross sections $\sigma_l^{(R)}(E)$ may be described by the simple function form

$$\sigma_l^{(R)}(E) = (2l+1)\left[1 + e^{\frac{(l-l_0)}{a}}\right]^{-1} + \varepsilon(2l_0 + 1)e^{\frac{(l-l_0)}{a}}\left[1 + e^{\frac{(l-l_0)}{a}}\right]^{-2} \qquad (3)$$

with the tabulated values of $l_0(E,A)$, $a(E,A)$, and $å(E,A)$ all varying smoothly with energy and mass. Those studies were initiated with the partial reaction cross sections determined by using complex, non-local, energy-dependent, optical potentials generated from a g-folding formalism (Amos *et al.* 2000). While those g-folding calculations did not always give excellent reproduction of the measured data (from ~ 20 to 300 MeV for which one may assume that the method of analysis is credible), they did show a pattern for the partial reaction cross sections that suggest the simple function form given in Eq. (3). With that form excellent reproduction of the proton total reaction cross sections for many targets and over a wide range of energies were found with parameter values that varied smoothly with energy and mass. We report now that the partial total cross sections for scattering of neutrons from nuclei can also be so expressed and we suggest forms, at least first average result forms, for the characteristic energy and mass variations of the three parameters involved. Nine nuclei, $^6$Li, $^{12}$C, $^{19}$F, $^{40}$Ca, $^{89}$Y, $^{184}$W, $^{197}$Au, $^{208}$Pb and $^{238}$U, for which a large set of experimental data exist, have been considered. Also those nuclei span essentially the whole range of target mass. However, to set up an appropriate simple function form, initial partial total cross sections must be defined by some method that is physically reasonable. Thereafter the measured total cross-section values themselves can be used to tune details, and of the parameter $l_0$ in particular. We chose to use results from *g*-folding optical potential calculations to give those starting values.

## III. RESULTS AND DISCUSSION

While we have used the partial total cross sections from DWA results for neutron scattering from all the nuclei chosen and at all of the energies indicated, only those obtained for $^{208}$Pb are shown in Fig. 1.

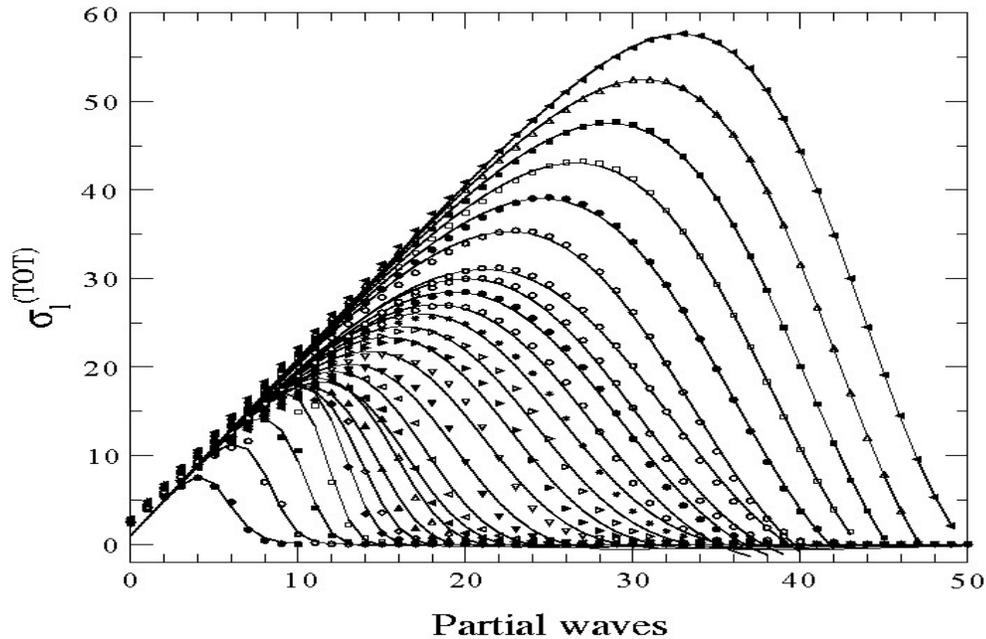

FIGURE 1. The partial total cross sections for scattering of neutrons from $^{208}$Pb for energies between 10 and 600 MeV. The target energy has the broadest spread of values.

The results from calculations of scattering from the other nuclei have similar form. The `data' shown as diverse open and closed symbols in Fig. 1 are the specific values found from the *g*-folding optical model calculations. Each curve shown therein is the result of a search for the best fit values of the three parameters, $l_0$, $a$, and $å$ that map Eq. (3) (now for total neutron cross sections) to these `data'. From the sets of values that result from that fitting process, the two parameters $a$ and $å$ can themselves be expressed by the parabolic functions

$$a = 1.29 + 0.00250E - 1.76 \times 10^{-6} E^2 \; ; \; \varepsilon = -1.47 - 0.00234E + 4.16 \times 10^{-6} E^2, \qquad (4)$$

where the target energy E is in MeV. There was no conclusive evidence for a mass variation of them. With $a$ and $å$ so fixed, we then adjusted the values of $l_0$ in each case so that actual measured neutron total cross-section data were fit using Eq. (3). The values of $l_0$ increase monotonically with both mass and energy and that is most evident in Fig. 2 where the optimal values $l_0(E)$ are presented as diverse filled or open symbols.

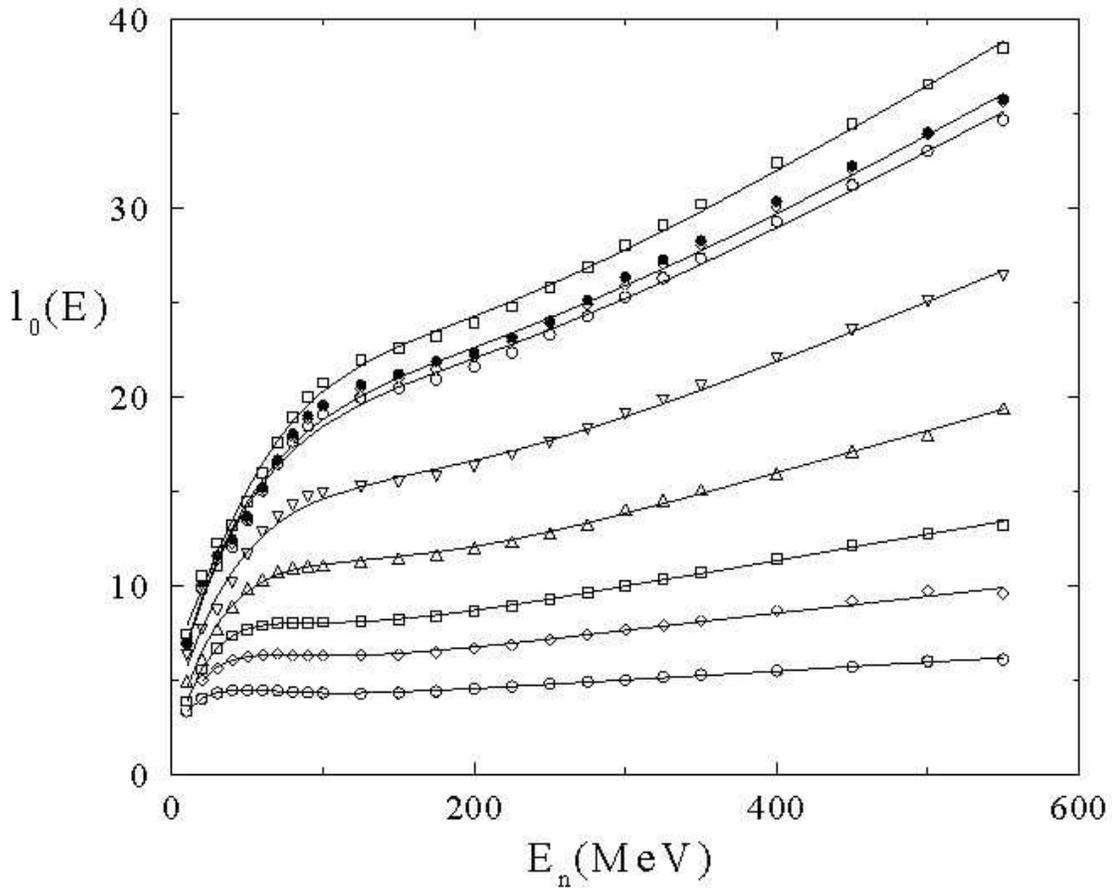

FIGURE 2. The values of $l_0$ that give best fits to the neutron total scattering cross section data. The curves are the best fit to these points assuming a function form for $l_0(E)$.

The set for each of the select nine masses (from 6 to 238) are given by those that increase in value respectively at 600 MeV. While that is obvious for most cases, note that there is some degree of overlap in the values for $^{197}$Au (opaque diamonds) and for $^{208}$Pb (filled circles). Plotting the values of $l_0$ against mass also reveals smooth trends as is evident in Fig. 3. Some actual energies are indicated by the numbers shown in this diagram. Again the curves shown in the figure are the results found on taking a functional form for $l_0 (A)$ at each energy (Deb and Amos, 2004).

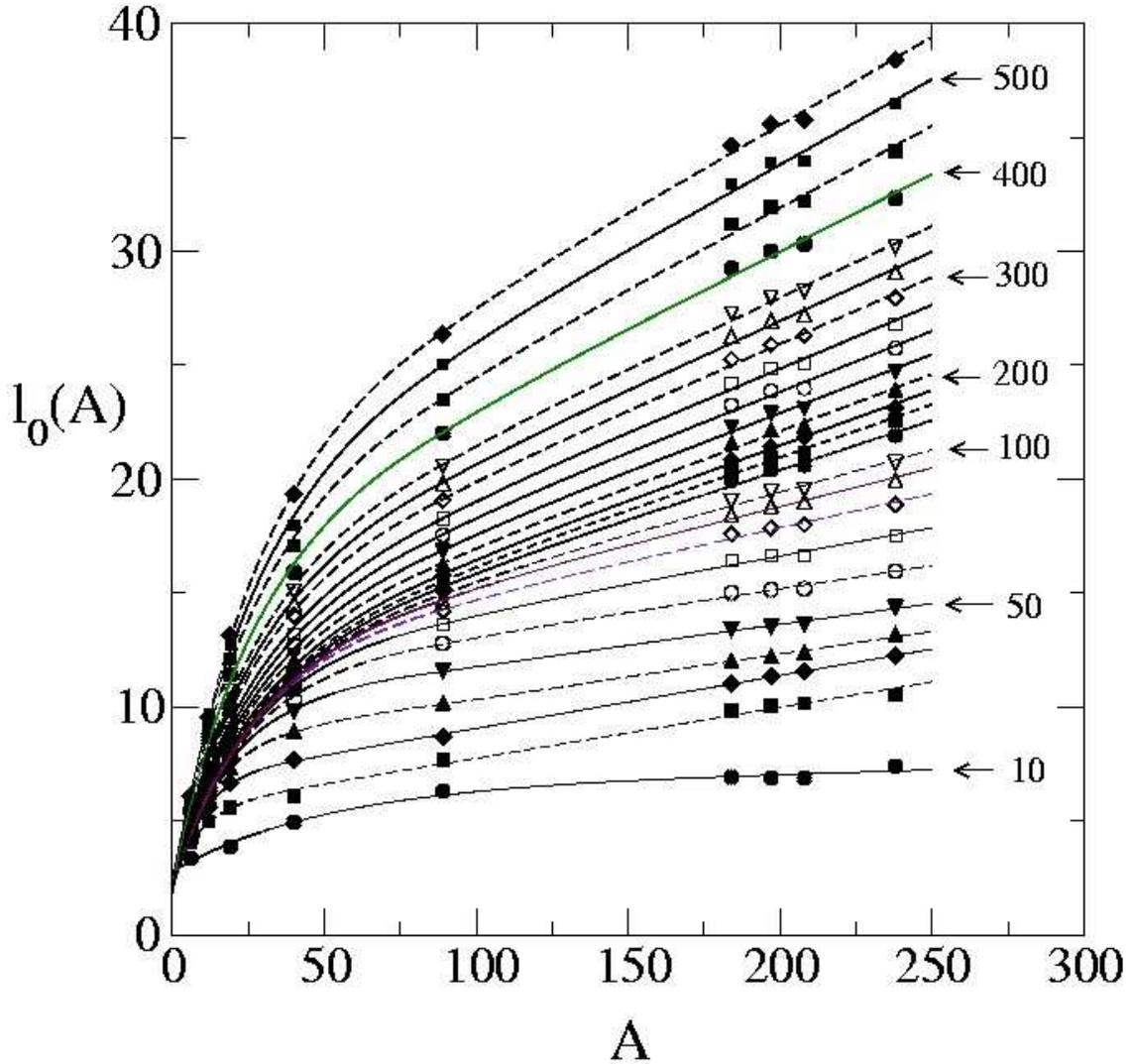

FIGURE 3. The values of $l_0$ depicted in Fig. 2 as they vary with mass for all of the energies considered. The curves are the best fit values for each mass assuming a function form for $l_0(A)$.

The total neutron scattering cross sections generated using the function form for partial total cross sections with the tabled values of $l_0$ (Deb and Amos, 2004) and the energy function forms of Eq. (4) for $a$ and $å$, are shown in Figs. 4-6. They are displayed by the continuous lines that closely match the data which are portrayed by opaque circles. The data that was taken from a survey by Abfalterer *et al.* (Abfalterer *et al.* 2001), which includes data measured at LANSCE that are supplementary and additional to those published earlier by Finlay *et al* (Finlay *et al.*, 1993). For comparison we show results obtained from calculations made using *g*-folding optical potentials (Amos *et al.* 2002). Dashed lines represent the predictions obtained from those microscopic optical potential calculations. Clearly for energies 300 MeV and higher, those predictions fail.

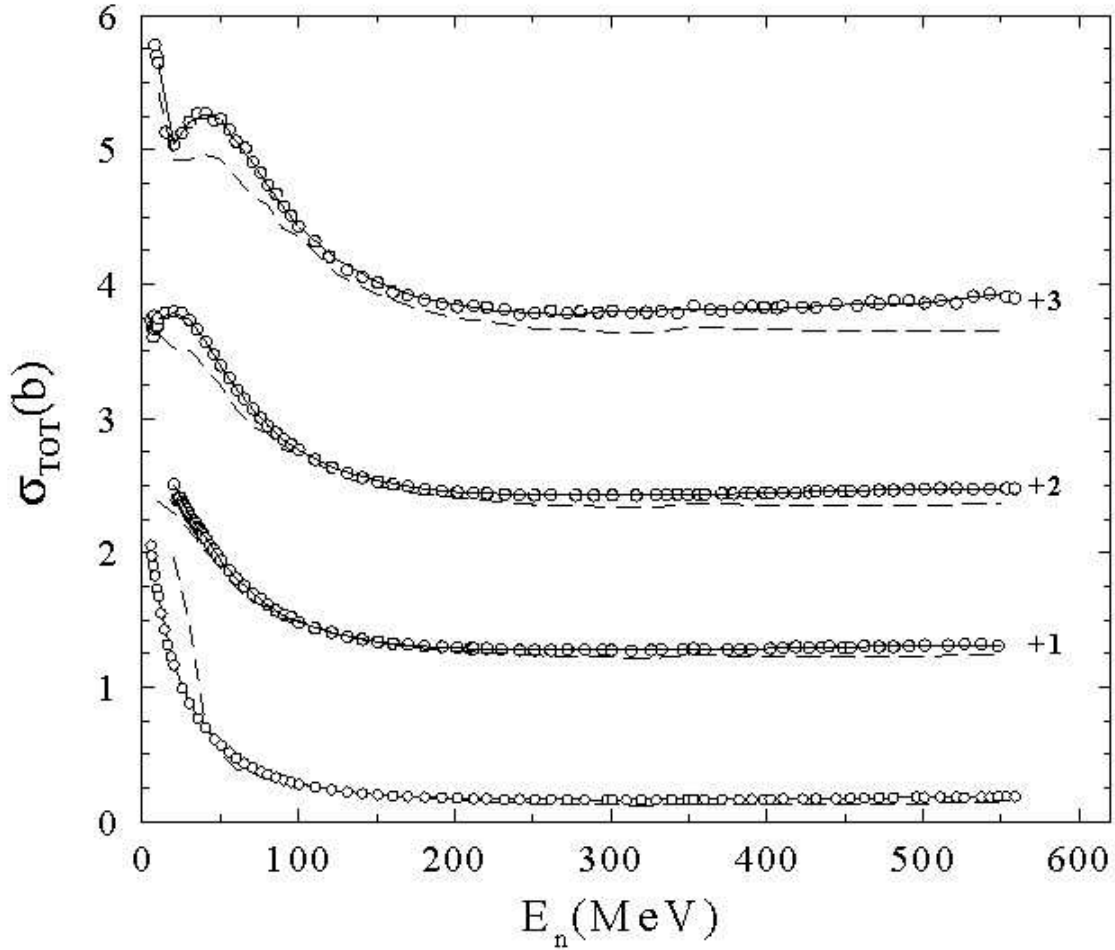

FIGURE 4. Total cross sections for neutrons scattered from $^6$Li, $^{12}$C, $^{19}$F, and $^{40}$Ca. The results have been scaled as described in the text to provide clarity.

The total cross sections for neutrons scattered from the four lightest nuclei considered are compared with data in Fig. 4. Therein from bottom to top are shown the results for $^6$Li, $^{12}$C, $^{19}$F, and $^{40}$Ca with shifts of 1, 2 and 3 b made for the latter three cases respectively to facilitate inspection of the four sets. A slightly different scaling is used in Fig. 4 in which the total neutron scattering cross sections from the nuclei $^{89}$Y (unscaled), $^{184}$W (unscaled), $^{197}$Au (shifted by 2 b), and $^{238}$U (shifted by 3 b) are compared with the base *g*-folding optical potential results and with the function forms with the optimal parameters. Again the *g*-folding potential results are displayed by the dashed curves while those of the function form are shown by the solid curves.

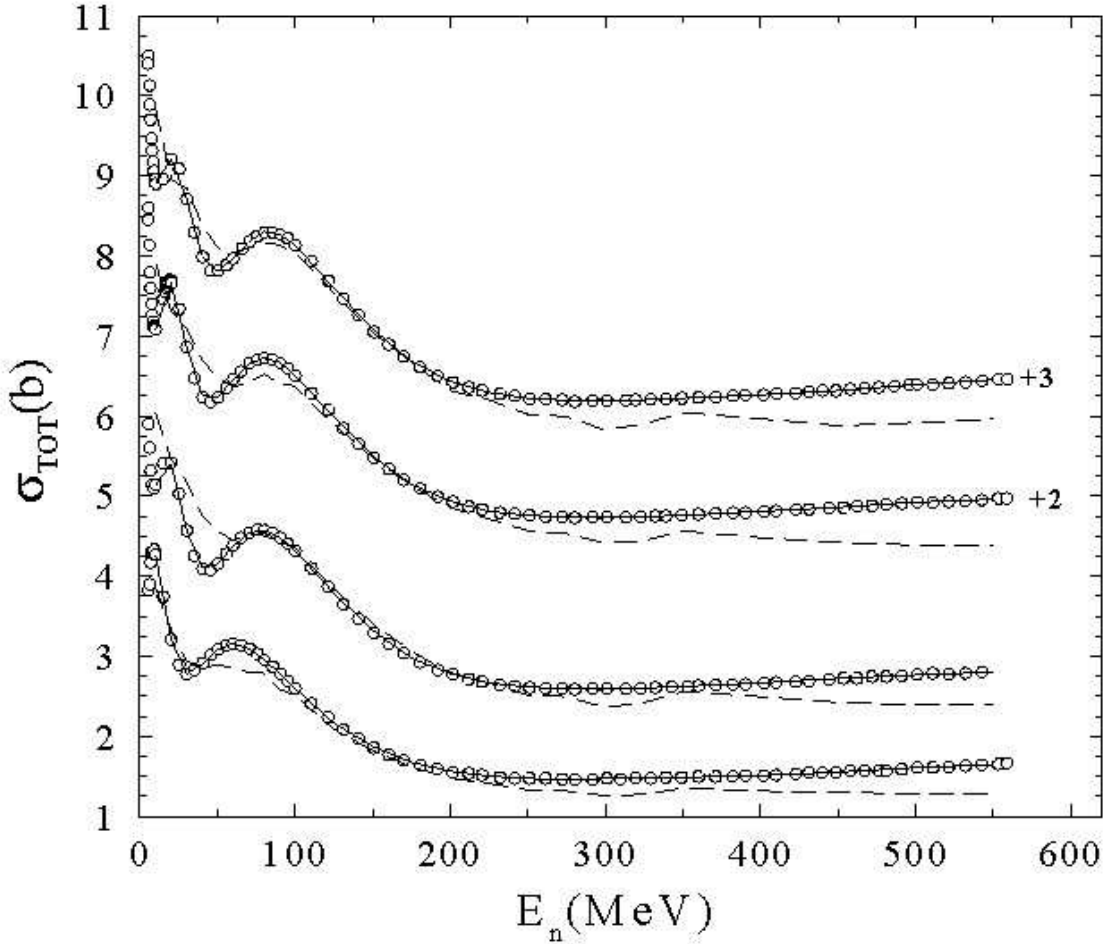

FIGURE 5. Total cross sections for neutrons scattered from $^{89}$Y, $^{184}$W, $^{197}$Au, and $^{238}$U. The results have been scaled as described in the text to provide clarity.

Finally we show in Fig. 6, the results for neutron scattering from $^{208}$Pb. In this case we used Skyrme-Hartree-Fock model (SKM*) densities (Brown 2000) to form the *g*-folding optical potentials. That structure when used to analyze proton and neutron scattering differential cross sections at 65 and 200 MeV gave quite excellent results (Karataglidis *et al.* 2002). Indeed those analyzes were able to show selectivity for that SKM* model of structure and for the neutron skin thickness of 0.17 fm that it proposed. Using the SKM* model structure, the *g*-folding optical potentials gave the total cross sections shown by the dashed curve in Fig. 6. Clearly there is a need to improve this model for energies at and above pion threshold. Nonetheless, it does do quite well for lower energies, most notably giving a reasonable account of the Ramsauer resonances (Koning and Delaroche, 2003) below 100 MeV.

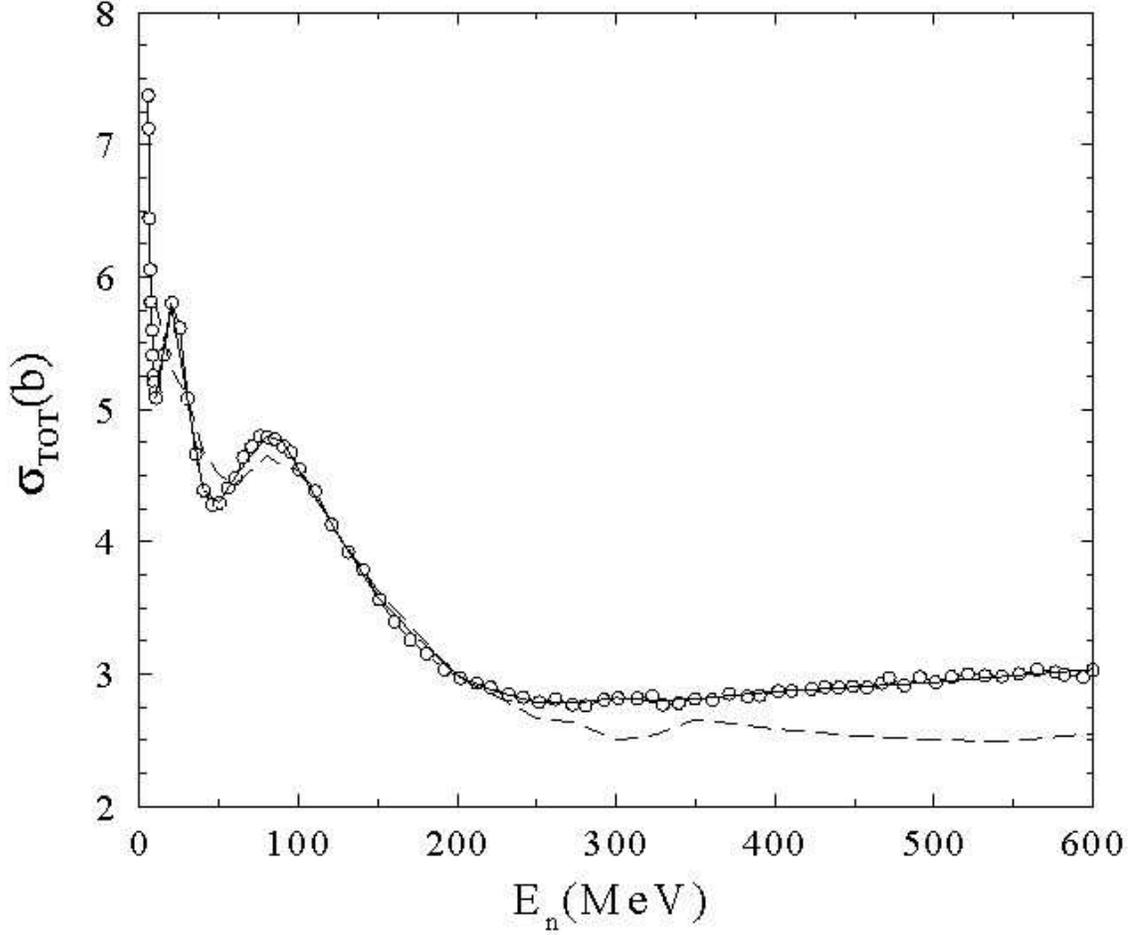

FIGURE 6. Total cross section for neutrons scattered from $^{208}$Pb.

## IV. RAMSAUER EFFECT

A Ramsauer-like effect has been included in past data analyses to describe the large scale variations of total cross sections from an otherwise smooth monotonic background (Dietrich *et al.* 2003). Under serious approximation, this correction is a coherent scaling of a theoretical model (Diffraction, global optical, and in our case, a functional form) of the background, namely, $\sigma_{TOT}(E, Data) = \sigma_{th}(E)[1 - \alpha(E)\cos(\beta(E))]$. Dietrich *et al.* (Dietrich *et al.* 2003) also linked this by Wick's limit to extract reaction cross sections. The validity of that was measured by the fractional deviation found for the zero degree cross section expectation and which is specified by (using $R(E) = á(E) \cos â (E)$),

$\eta(\%) = 100\left(\dfrac{\alpha(E)\sin(\beta(E))}{R(E)}\right)^2$. Using these, the optimal background parameters yield a

Ramsauer effect (Deb *et al.*, 2004) that continues to high energies but which has a very regular character (when results are displayed in logarithmic-linear plots). The two parameters involved vary very smoothly, and result in tolerably small variations from Wick's limit. That latter variation in fact has half the wavelength of the Ramsauer effect in the data.

## V. CONCLUSIONS

We suggest a three parameter function form for partial total cross sections that will give neutron total cross sections without recourse to phenomenological optical potential parameter searches. That functional form also reproduces proton reaction cross sections. The parameters that fit actual data show smooth trends with both energy and target mass. Taking energy dependent forms for them with a Ramsauer model correction allowed replication of the total cross section for neutron-$^{208}$Pb scattering for energies from 5 to 600 MeV. The scheme also involves minimal deviation from Wick's limit.

**References**
Abfalterer, W. P. *et al*. (2001). *Phys. Rev.* C **63**, 044608.
Amos, K., *et al*. (2000). *Ad. Nucl. Phys*. **25**, 275.
Amos, K., *et al*. (2002). *Phys. Rev.* C **65**, 064618.
Amos, K., *et al*. (2002). *J. Nucl. Sci. Technol., Suppl.* **2**, 738.
Brown, B. A., (2000). *Phys. Rev. Lett*. **85**, 5296.
Deb, P. K. and Amos, K. (2003). *Phys. Rev.* C **67**, 067602.
Deb, P. K. and Amos, K. (2004). *Phys. Rev.* C **69**, 064608.
Deb, P. K. *et al*. (2004). *Phys. Rev.* C **70**, 057601.
Dietrich, *et al*. (2003). *Phys. Rev.* C **68**, 064608.
Finlay, R. W., *et al*. (1993). *Phys. Rev.* C **47**, 237.
Karataglidis, S., *et al*. (2002). *Phys. Rev.* C **65**, 044306.
Koning, A. J., and Delaroche, J. P. (2003). *Nucl. Phys.* A **713**, 231.